\documentclass[prl,twocolumn,showpacs,preprintnumbers,amsmath,amssymb]{revtex4}

\usepackage{graphicx}
\usepackage{dcolumn}
\usepackage{bm}

\begin{document}


\title{Secondary electron cascade in attosecond photoelectron spectroscopy from metals}

\author{Jan Conrad Baggesen}
 \affiliation{Lundbeck Foundation Theoretical Center for Quantum System Research \\
Department of Physics and Astronomy, Aarhus University, 8000 Aarhus C, Denmark.}

\author{Lars Bojer Madsen}%
\affiliation{Lundbeck Foundation Theoretical Center for Quantum System Research \\
Department of Physics and Astronomy, Aarhus University, 8000 Aarhus C, Denmark.}

\date{\today}


\begin{abstract}
Attosecond spectroscopy 
is currently restricted to photon energies around 100 eV. We show that under these conditions, electron-electron scatterings, as the photoelectrons leave the metal give rise to a tail of secondary electrons with lower energies and hence a significant background. We develop an analytical model based on an approximate solution to Boltzmann's transport equation, to account for the amount and energy distribution of these secondary electrons. Our theory is in good agreement with the electron spectrum found in a recent attosecond streaking experiment. To suppress the background and gain higher energy resolution, photon sources of higher energy could be advantageous.
\end{abstract}

\pacs{42.50.Hz, 42.65.Re, 79.60.-i, }
\maketitle

The time-resolved study of metal surfaces using femto- and attosecond pulses is currently developing as a separate research topic~\cite{Apolonski2004,Miaja-avila2006,Cavalieri2007,miaja-avila2008,Saathoff2008,Miaja-avila2009}. It is now possible to perform experiments with a very high temporal resolution, allowing time-resolved investigations of electron dynamics. These include collective dynamics such as plasmon creations~\cite{Kubo2005,Stockman2007}, electronic screening within the metal~\cite{Huber2001} and image-charge states~\cite{Wolf1997}. When combining XUV pumps and infrared probes, it is possible to study in real time the Auger decay from atoms adsorbed on a metal surface~\cite{miaja-avila2008} and using the streaking technique~\cite{Goulielmakis2004}, to resolve a delay of only 110 asec between photoelectrons emerging from the conduction band and core levels~\cite{Cavalieri2007}.

A series of theory papers study the short-pulse interactions with metal surfaces~\cite{Lemell03,Baggesen2008,Zhang2009,Freericks2009,Kazansky2009}.
Progress in this field is challenging since field-free, solid state methods are not readily extended to account for the highly nonlinear and explicitly time-dependent dynamics introduced by  the short pulses.  Even when it is possible to regard the interaction in a one-electron model, many different secondary processes may blur the direct process.
Hence, development of models that capture the essential physics and are still computationally tractable are called for. Here, we develop a simple, analytical description of the electron-electron scattering in the metal and show that this leads to a large background of electrons $\sim E^{-2}$. This energy-dependent background is a result of the spectral broadness of the attosecond pulse and leads to a reduced energy resolution for any signal from electrons below the conduction band. To suppress the background, high-energy photon sources should be used.

Our theory is based on the fact that fast primary electrons created by the attosecond pulse, will 
produce secondary electrons  by collisions with electrons bound in the metal, and  initiate a cascade as they move  through the metal. The primaries and secondaries lose energy until they escape, if sufficiently energetic, or sink back into the Fermi sea. The secondaries are responsible for the low-energy tail in the spectrum. To find the spectrum, including the above mechanism, implies the solution of a full dynamic many-body problem. Instead of persuing a quantum solution, we aim at a semi-analytical approach, where the energy distibution is determined by the solution of a transport equation for the electrons. To formulate the transport  equation, we use that the main way the electron loses energy in the energy range of interest is by collision with conduction electrons;
the electron-phonon coupling is weaker~\cite{Wolff54}.
In a steady-state regime, and based on neutron transport theory~\cite{Marshak47}, casade processes pertaining to this situation were considered many years ago for an electron beam impinging normally to a metal surface~\cite{Wolff54}.  In the present scenario, however, the pulsed light sources introduce an explicit time-dependence in the problem, and our starting point is the time-dependent Boltzmann  equation~\cite{Reif}
\begin{align}
&	\left[\frac{\partial }{\partial t} + {\bm v} \cdot \frac{ \partial }{\partial {\bm r}}  + \frac{{\bm F}(t)}{m_e}  \cdot \frac{ \partial }{\partial {\bm v}} \right ]  \Phi ({\bm r}, {\bm \Omega}, E, t) = 
\nonumber \\
&-\frac{v\Phi({\bm r}, {\bm \Omega}, E, t)}{\lambda(E)}
+ J({\bm r}, {\bm \Omega}, E, t) 
\nonumber \\
&+ \int_E^\infty dE'\int d{\bm\Omega}' \frac{v'\Phi({\bm r}, {\bm \Omega}', E', t )}{\lambda(E')} S(E, {\bm \Omega}; E', {\bm \Omega}'),
	\label{eq:Boltzmann}
\end{align}
where $\Phi ({\bm r}, {\bm \Omega}, E, t) d{\bm r} d {\bm \Omega} dE $ is the electron distribution function, i.e., the number of electrons in $[{\bm r}; {\bm r} + d {\bm r}], [{\bm \Omega}; {\bm \Omega} +d {\bm \Omega}], [E; E + dE]$ at time $t$, ${\bm r}$ is the position within the metal, ${\bm \Omega} = {\bm v}/v$ is a unit vector in the direction of the electron velocity $v$, and $E$ is the energy. The LHS in \eqref{eq:Boltzmann}, 
represents the time-rate of change of the electron distribution function in the direction ${\bm \Omega}$, including the effect of the force ${\bm F}(t)$ from the streaking near-ir fs  pulse. On the RHS the first term describes the loss of electrons due to scattering and possibly capture, and $\lambda(E)$ is the mean free path. $J$ is the source term resulting from the attosecond pulse, giving the number of electrons emitted per unit time at ${\bm r}$ and $t$ with energy $E$ in direction  ${\bm \Omega}$. The factor ${v'\Phi({\bm r}, {\bm \Omega}', E', t )} / {\lambda(E')}$ in the last term describes the number of collisions per unit time at ${\bm r}$ occuring to electrons with energy $E'$ and direction ${\bm \Omega}'$, finally, $S(E, {\bm \Omega}; E, {\bm \Omega}')$
is the probability that, given an electron at $E', {\bm \Omega}'$, one will be found at $E, {\bm \Omega}$ after a scattering.

First, we disregard the fs, ir-pulse and focus 
on the influence of electron propagation on the number of electrons passing per unit surface area induced by the attosecond uv pulse alone. We therefore leave out the force term and integrate \eqref{eq:Boltzmann} over the attosecond pulse and use that no contribution from the first term on the LHS of \eqref{eq:Boltzmann} appears since no continuum electrons are present inside the metal long before or after the pulse. The  equation describing the transport and energy distribution of electrons as a consequence of  the attosecond pulse accordingly reads
\begin{align}
\lambda(E) & ({\bm \Omega} \cdot \nabla) N({\bm r},  {\bm \Omega}, E) + N({\bm r},  {\bm \Omega}, E) 
=
\lambda(E) P({\bm r}, E, {\bm \Omega})  \nonumber \\ +& \lambda(E) \int_E^\infty \int d {\bm \Omega}' \frac{N({\bm r},{\bm \Omega}', E') }{\lambda(E')} S(E, {\bm \Omega} ; E', {\bm \Omega}'),
\label{eq:Boltzmann2}
\end{align}
where $N({\bm r}, {\bm \Omega}, E) = \int_{-\infty}^{\infty}  dt v \Phi$ is the number of electrons in point ${\bm r}$ passing unit area along ${\bm \Omega}$ with energy $E$, and $P({\bm r}, {\bm \Omega}, E) =  \int_{-\infty}^{\infty}  dt J$ is the number of source electrons originating from a single pulse. 
Equation \eqref{eq:Boltzmann2} is still too complicated to be solved analytical, so to proceed we introduce a series of approximations.
 The mean free path $\lambda(E)$ is $\sim 5$~\AA~ in the entire range 30 - 100 eV~\cite{Zangwill}, and is taken as a constant $\lambda= 5$~\AA~ in the following. The electron distribution within the metal is assumed to depend only on the distance to the surface, $\vert z \vert$. The skin-depth $\delta_X$  for the uv radiation is several  nm's even at grazing incidence, i.e., $\delta_X \gg \lambda(E)$. Accordingly, the spatial derivative in \eqref{eq:Boltzmann} is also neglected, and the source term $P$ is independent of $z$. The problem is then formally identical with the electron beam case considered previous, and we follow~\cite{Wolff54} and 
 expand $N$, $S$ and $P$ in Legendre polynomials in the angle between the velocity of the secondary electron and the surface normal.  We use that for electron energies up to $\sim 100$ eV the spherical part dominates, i.e., keep only the $\ell = 0$ component in the Legendre expansion, and 
we then take the $s$-wave result $ S_0 = 2 /E'$~\cite{Wolff54}, where the factor of $2$ takes into account that electrons may appear at energy $E$ either by scattering or by transitions from the conduction band, and where the subscript denotes $\ell = 0$. The corresponding Boltzmann equation reads
$N_0(E) = \lambda P_0(E)  + \int_E^\infty d E' N_0(E') \frac{2}{E'}$,
and as is readily seen by insertion, the solution 
is
$	N_0(E) = - \lambda \int_E^\infty dE' \Big(\frac{E'}{E}\Big)^2\frac{\partial P_0}{\partial E'}.$ 
We integrate 
this expression  by parts, and find that the number of electrons at energy $E$ per attosecond pulse, i.e., the photoelectron spectrum,  is given by
\begin{equation}
	N_0(E) = \lambda P_0(E) + 2\int_E^\infty dE'\frac{E'}{E^2} \lambda P_0(E').
	\label{eq:cascade}
\end{equation}
 The first term in \eqref{eq:cascade},  describes direct electrons and the second secondary cascade electrons.

To correct for the finite probability of escaping through the surface, we multiply $N_0(E)$ from~\eqref{eq:cascade} by the normal component of the free-electron transmission probability~\cite{Merzbacher} $T(E) = 4\sqrt{1-\frac{V_0}{E}}/{\Big(1+\sqrt{1-\frac{V_0}{E}}\Big)^2}$, with $V_0$ the sum of the work function and the Fermi energy, $E_F$, of the metal.

To evaluate the spectrum, we see from \eqref{eq:cascade} that we need  the number of electrons  $\lambda P_0(E)$ created with energy $E$, within the distance $\lambda $ of the surface.  The effects of electron-electron scattering will not be too sensitive to the exact form of $\lambda P_0(E)$, so for convenience we follow~\cite{Baggesen2008,Zhang2009}, and consider a $T$-matrix formalism for a jellium-like metal. In this model we take $\lambda P_0   = \sum_\text{Occupied states}|T_{fi}|^2\rho(E_f),$
where the $T$-matrix element for the uv-driven transition 
from the initial $\vert \Psi_i(t) \rangle$ and to the final state $\vert \Psi_f(t) \rangle$ may be written as
$	T_{fi} = -i\int_{-\infty}^{\infty}dt\langle\Psi_f(t)|V_{X}(t)|\Psi_i(t)\rangle$ and $E_f$ is the energy of the final state.
The interaction is 
$	V_X(t) = \vec r \cdot \vec \epsilon E_X(t)$,
with $\vec \epsilon$ the polarization vector and $E_X(t)$ the field strength of the uv pulse. We  include conduction  and localized core electrons. The latter are modeled  by
$	\Psi_{loc}^{\vec{k}}(\vec r) = \frac{1}{\sqrt{N}}\sum_{\vec{R}}e^{i\vec k \cdot \vec R} \psi_{loc}(\vec r - \vec R)$,
where $\vec R$ runs over all atom positions within the material, $\psi_{loc}(\vec r)$ is the atomic localized state and $N$ is the total number of atoms in the metal. To model the surface, it is required to make linear combinations of these solutions, which vanish at the surface, $z=0$. This is fulfilled by taking the initial localized states as
$	\Psi_{loc}^{\vec{k}}(\vec r) = \psi_{loc}^{\vec{k}}(\vec r) - \psi_{loc}^{-\vec{k}}(\vec r)$
inside the metal. Outside the metal, the wave function is vanishing.

The conduction electrons are modeled with free-electron states~\cite{Baggesen2008,Zhang2009}. Inside the metal, $\Psi_{cb}^{\vec k} = \frac{e^{i\vec k_{||}\cdot \vec r_{||}}}{(2\pi)^{3/2}\sqrt{V}}\Big(e^{ik_z z} + \frac{k_z -i\gamma}{k_z + i\gamma}e^{-ik_z z}\Big)$,
where $V$ is the total volume of the metal and $\gamma = \sqrt{2V_0 - k^2}$, $k^2 / 2 < E_F$. 

For the final states, we use  free-electron states,  damped within the metal to account for the mean free path, i.e.,
$	\Psi_f(\vec{r},t) =  \frac{1}{\sqrt{V}}e^{-\frac{r}{2\lambda}} e^{i \vec{k}_f \cdot \vec{r}} e^{- i \frac{{k}_f^2 t}{2}}$. The density of states needed to perform the integration over the continuous band of $k$ points within the solid, is with the current choice of normalization
$	\rho(\vec k) d^3k = \frac{V}{(2\pi)^3}d^3k$,
for both conduction and core electrons~\cite{AshcroftMermin}. When comparing the amount of electrons released from the conduction band and the core states to experimental results\cite{Cavalieri2007}, the contribution from the delocalized conduction band, which is not an accurate description of tungsten\cite{Christensen1974}, comes out an order of magnitude too small. The relative contribution has been scaled to represent the experiment.

We apply our  theory to tungsten and take experimental parameters~\cite{Cavalieri2007}: We use a gaussian envelope for the uv pulse, with $300\text{ asec}$ intensity-FWHM  and a central frequency 91 eV. 
 We use an experimental Fermi energy of 4.5 eV for the conduction band, a hydrogenic localized 1s-state with a binding energy of 32.5 eV and a tungsten work function of 5.5 eV. Note that the series of experiments~\cite{Miaja-avila2009,miaja-avila2008,Saathoff2008,Miaja-avila2006} apply an analysis based on laser-induced side bands and focus on energies so close to the main uv peaks, that  we have not been able to assess the influence of the background in those data.
\begin{figure}
\includegraphics[width = 0.45\textwidth]{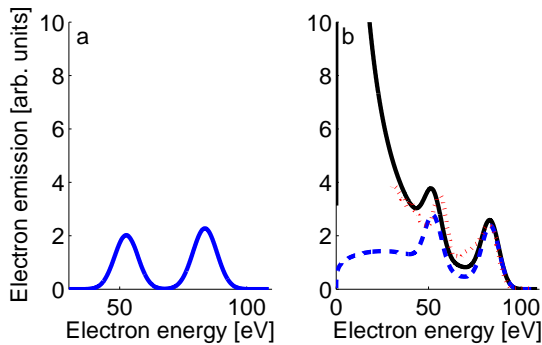}
	\caption{(Color online) Calculation of the photoelectron spectrum for a 300 asec uv pulse, including (a) only the direct electrons and including (b) both direct and scattered electrons. The full (black) curve is with the cascade of secondary electrons while the dashed (blue) curve includes only one scattering. The contribution from the conduction band has been scaled to be comparable to the contribution from the core band. The dotted curve in (b) show data from~\cite{Cavalieri2007}.}
	\label{fig:fig1}
\end{figure}

In Fig.~\ref{fig:fig1}, we consider the spectra without assisting ir field to clearly display typical spectra of (a) direct electrons and (b) secondary electrons that have undergone scattering. Figure \ref{fig:fig1}(a) shows that direct emission leads to two peaks, representing electrons released from the deeper-lying localized states and from the conduction band. When the electron-electron scattering is included, the two peaks both have a tail of lower energy electrons. This situation is shown in Fig.~\ref{fig:fig1}(b) where we also show the experimental results without ir field~\cite{Cavalieri2007}. We see that the inclusion of a cascade of electron scatterings explains the background in the spectrum. As an alternative to the cascade theory, we have calculated the electron distribution including direct electrons and electrons that has suffered exactly one scattering, similarly to \cite{Berglund1964}. This reproduces the amount of secondaries with energy near the highest energy direct peak, but fails to account for the very many low-energy electrons that are released, supporting that multiple scatterings and a cascade is at play. Note that the experimental data have been subtracted for electrons stemming from above threshold ionization (ATI)~\cite{Cavalieri2007}, and hence these do not contribute to the background.   

Due to space-charge effects not included in our model, the experimental peaks are expected to be upshifted by approximately 3 eV~\cite{Cavalieri2007}. While this shift is seen for the core-level electrons in Fig.~\ref{fig:fig1}(b), the experimental and theoretical peaks pertaining to the conduction  electrons coincide. This accidental agreement is associated with a short-coming of the free electron model which captures the qualitative features but  does not  quantitatively describe the tungsten conduction band, which holds a lot of structure~\cite{Christensen1974}. Consequently, 
the center of the peak comes out of our calculation a few eV too high, since many of the conduction band electrons are actually a few eV deeper bound~\cite{Christensen1974}.

\begin{figure}
\includegraphics[width = 0.45\textwidth]{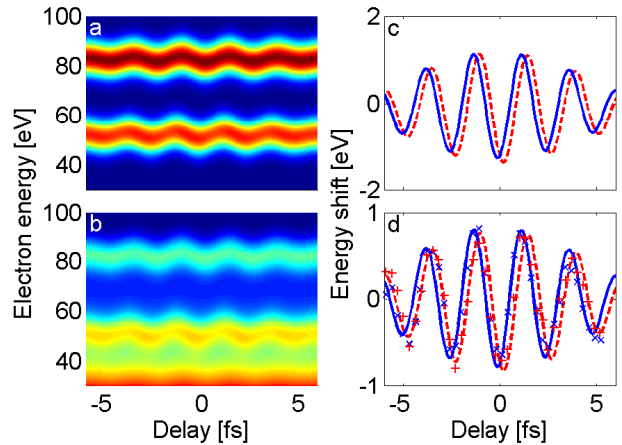}	
	\caption{(Color online) Results for tungsten, with a laser intensity $5\times10^{10}\text{ W/cm}^2$. (a) Direct emission spectra with scaled contribution from the conduction band. (b) Emission spectra including scattering to lower energies. (c) Centre-of-energy analysis showing the temporal delay of the 4f-electrons (full, blue) relative to the conduction band electrons (dashed, red) for only the direct electrons. (d) The same as (c) but including scattering. Also shown are experimental data from~\cite{Cavalieri2007};  cross: 4f electrons; plus: conduction band electrons. The energies of the 4f-electrons has been multiplied by 1.1 in (c) and by 1.8 in (d). 
	Notice the difference is scale in (c) and (d). See text for laser parameters.}
	\label{fig:fig2}
\end{figure}

We now turn to the streaking spectra. We follow~\cite{Zhang2009}, and invoke the streaking directly in the source term by adding the appropriate Volkov phase to the final free-electron states. For the ir pulse we use a sine-squared envelope, a duration $13\text{ fs}$, a wavelength  of 750 nm, and an intensity of   $5 \times 10^{10}$ W/cm$^2$. Comparing Figs.~\ref{fig:fig2}(a) and (b) shows the effect of including electron-electron scattering as the primary photoelectron produced by the attosecond uv pulse propagates to the surface and produces a cascade of secondary electrons. We see that the structure and magnitude of the background mathes quite well that seen in experiment after ATI electrons has been subtracted (Fig.~2(b) in \cite{Cavalieri2007}).

Figures \ref{fig:fig2}(c) and (d) show the corresponding centre-of-energy (CoE) spectra.
The CoE analysis is performed on the  interval 44 eV - 63 eV (4f-electrons) and above 66 eV (conduction band electrons).  The 4f-electron curve is, however, on top of the conduction band electron tail and is damped more than the conduction band curve. As in \cite{Zhang2009}, the 
4f curve has been multiplied by a factor (in our case 1.8) to make the 4f and conduction band amplitudes comparable. 
We see from a comparison of Figs.~\ref{fig:fig2}(c) and (d) that the inclusion of the background produced by secondary electrons leads to a reduction of the amplitude in the CoE spectrum and a very good  agreement with the measured spectrum~\cite{Cavalieri2007}. We note that the amplitude in the CoE data can be reproduced  with the present $T$-matrix theory not including the background, but for an unrealistic  low  intensity of the femtosecond laser pulse. In this respect it is essential to have access to data as in Fig.~\ref{fig:fig2}(b).

Although the issue of the time-delay is not a prime concern in the present work, we make the following comment.  
We find a shift of $\sim 300$ asec which is larger than what was found experimentally~\cite{Cavalieri2007} and in a similar model~\cite{Zhang2009}. 
This discrepancy might be due to the failure of the jellium model~\cite{Baggesen2008,Zhang2009} to describe the structured conduction band in tungsten~\cite{Christensen1974}, and points to the need for further work to fully understand the origin of the time delay, e.g., along the lines of time-dependent wavepacket calculations~\cite{Kazansky2009}. In this connection, we also note that in~\cite{Baggesen2008,Zhang2009} and this work, the streaking takes place inside the metal assuming the penetration depth of the ir field to be much larger than the mean free path, while in~\cite{Kazansky2009}, the streaking occurs as the electron wave packet leaves the surface. 

With the inclusion of electron scattering within the metal, we are able to reconstruct the significant tail of electrons at lower energies observed experimentally in attosecond photoelectron spectroscopy from metals~\cite{Cavalieri2007}. In monochromatic experiments there is generally relatively  few electrons scattered to lower energies, as compared to what is seen  in the attosecond regime. This is due to the fact that the amount of electrons scattered is proportional to the total amount of excited electrons. In an experiment with a spectrally broad pulse, the tail of electrons becomes comparable to the main peak of direct electrons. 

In conclusion, we have presented a simple model for electron scattering effects in the attosecond domain for electron propagation through a metal, and we have shown that the model reproduces two sofar unexplained central features of the spectra in this regime (i) the background and (ii) the amplitude in the  streaking spectrum.
Electron-electron scattering is an important process to consider to fully understand attophysics in the condensed phase. The significance of the secondary electrons relative to the primary electrons is much greater in attosecond spectroscopy than in conventional photoelectron spectroscopy due to the broad bandwidth of the pulse. Our findings \eqref{eq:cascade} show that the relative amount of secondary electrons near the direct peak scales as $E^{-1}$. Hence, to obtain higher energy resolution, it would be advantageous to use higher photon energies as may become possible with the upcoming free-electron laser sources. The present work shows that the classical Boltzmann equation captures important physics in the considered regime and, hence, in combination with input from quantum mechanical calculations,  provides an attractive starting point for further theory development.

After the submission of this manuscript, another work using classical transport theory to explain attosecond streaking appeared\cite{Lemell2009}.

This work was supported by the Danish Research Agency (Grant no. 2117-05-0081)


\end{document}